\documentclass[useAMS,usenatbib,usegraphicx]{mn2e}
\usepackage{times}
\usepackage{amsmath}
\usepackage{amssymb}

\usepackage{color}
\usepackage{url}
\usepackage{lscape}
\usepackage{rotating}
\usepackage{amssymb,subfigure,gensymb}

 \newcommand{\apj}{ApJ}
\newcommand{\apjl}{ApJL} \newcommand{\mnras}{MNRAS}
\newcommand{\aj}{AJ} \newcommand{\apjs}{ApJS}
\newcommand{\nat}{Nature} 
\newcommand{\aap}{A\&A} 
 \newcommand{\actaa}{Acta Astron.}
\newcommand{\apss}{APSS} \newcommand{\inprep}{\emph{in prep}}   

\begin{document}

\title[J band Variability of M Dwarfs in the WFCAM Transit Survey]{J
band Variability of M Dwarfs in the WFCAM Transit Survey}

\author[Goulding et al.] {N. T. Goulding$^1$, J. R. Barnes$^1$,
D. J. Pinfield$^1$, G. Kov\'{a}cs$^2$, J. Birkby$^3$, S. Hodgkin$^2$,
\newauthor S. Catal\'{a}n$^1$, B. Sip\H{o}cz$^1$, H. R. A. Jones$^1$,
C. del Burgo$^{4,5}$, S. V. Jeffers$^6$, S. Nefs$^3$, \newauthor
M.-C. G\'{a}lvez-Ortiz$^{1,7}$, E. L. Martin$^7$ \\ \\ \footnotesize
$^1$ Centre for Astrophysics Research, Science \& Technology Research
Institute, University of Hertfordshire, Hatfield, Herts, AL10 9AB, UK
\\ \footnotesize $^2$ Institute of Astronomy, University of Cambridge,
Madingley Road, Cambridge, CB3 0HA, UK \\ \footnotesize $^3$ Leiden
Observatory, Universiteit Leiden. P.O. Box 9513, 2300 RA Leiden,
Netherlands \\ \footnotesize $^4$ Instituto Nacional de Astrof\'{i}sica, 
\'{O}ptica y Electr\'{o}nica (INAOE), Aptdo. Postal 51 y 216, 72000 
Puebla, Pue., Mexico \\ \footnotesize $^5$ UNINOVA-CA3, Capus da Caparica,
2825-149 Caparica, Portugal  \\ \footnotesize $^6$ Institut f\"{u}r
Astrophysik, Friedrich-Hund-Platz 1, 37077, G\"{o}ttingen, Germany \\
\footnotesize $^7$ Centro de Astrobiolog\'{i}a (CSIC-INTA), Crta de
Ajalvil km 4. E-28850 Torrej\'{o}n de Ardoz, Madrid, Spain 
Somewhere Else \\}

\maketitle

\begin{abstract} We present an analysis of the photometric variability
of M dwarfs in the WFCAM Transit Survey. Although periodic lightcurve
variability in low mass stars is generally dominated by photospheric
star spot activity, M dwarf variability in the $J$ band has not been
as thoroughly investigated as at visible wavelengths. Spectral type
estimates for a sample of over 200,000 objects are made using spectral
type-colour relations, and over 9600 dwarfs ($J$$<$17) with spectral
types later than K7 were found. The light curves of the late-type
sample are searched for periodicity using a Lomb-Scargle periodogram
analysis. A total of 68 periodic variable M dwarfs are found in the
sample with periods ranging from 0.16 days to 90.33 days, with
amplitudes in the range of $\sim0.009$ to $\sim0.115$ in the $J$
band. We simulate active M dwarfs with a range of latitude-independent
spot coverages and estimate a periodically variable fractions of 1-3
per cent for stars where spots cover more than 10 per cent of the
star's surface. Our simulated spot distributions indicate that
operating in the $J$ band, where spot contrast ratios are minimised,
enables variability in only the most active of stars to be
detected. These findings affirm the benefits of using the $J$ band for
planetary transit searches compared to visible bands. We also
serendipitously find a $\Delta$$J$$>$$0.2$ mag flaring event from an
M4V star in our sample.\\\\
\end{abstract}


\section{Introduction}\label{sec:Introduction} The number of exoplanet
discoveries has been growing at an increasing rate, while the lower
mass limit for habitable-zone planets grows increasingly closer to
1M$_{\oplus}$. Indeed, planets of less than 1M$_{\oplus}$ orbiting
sun-like stars have been recently been detected
\citep{Fressin2011}. Recent discoveries by \citet{Vogt2010} and
\citet{Charbonneau2009} have highlighted how rich in low mass planets
M dwarf systems can be.  While radial velocity surveys have been
productive in detecting exoplanet M dwarf systems, only seven of the
251 known transiting planets have been detected orbiting M
dwarfs\footnote{exoplanet.eu database}, in four systems: GJ 436
\citep{Coughlin2008,Gillon2007}, GJ 1214 \citep{Charbonneau2009}, KOI
961 \citep{Muirhead2012}, and KOI-254 \citep{Johnson2012}. The Kepler
satellite has so far identified more then 30 early M dwarf planetary
host star candidates from 2510 stars cooler than 4000K being monitored
\citep{Borucki2011}. The Wide Field Camera (WFCAM) Transit Survey
(WTS) is a UK Infra-red Telescope (UKIRT) Campaign Survey comprising
more than 100 nights worth of observations. Its main science driver is
to identify planets orbiting M dwarf stars. The observations comprise
four fields, each of which is itself comprised of eight pointings of
the of WFCAM, which uses four 2048$\times$2048 pixel imaging arrays
each covering 13.65'$\times$13.65'. The most complete field has
accumulated a total of 1197 epochs of observations over 5 years. Light
curves were obtained in the $J$ band only with single deep exposures
in $ZYJFK$ bands. A full description of the observation technique is
given by \citet{Nefs2012}. Data reduction and light curve production
has been carried out by the Cambridge Astronomical Survey Unit (CASU)
using a customised pipeline \citep{Irwin2007,Kovacs2012}. In addition
to transit searches the WTS lightcurves facilitate a wide range of
additional science, including the search for and study of variable M
dwarfs, on which our work is focused. The sampling and overall time
base-line of the survey potentially permits sensitivity to periods up
to tens of days.

\begin{table}
\begin{tabular}{ | l | l | l | l | l | l | l | l | l | } \hline Name &
Coordinates & No. of epochs & Objects (J$<$17) \\ & RA (h), Dec (deg)
& & \\ \hline 19 & 19.58+36.44 & 1197 & 69161 \\ 17 & 17.25+03.74 &
832 & 17103 \\ 07 & 07.09+12.94 & 770 & 21224 \\ 03 & 03.65+39.23 &
589 & 15159 \\ \hline
\end{tabular}
\caption{A table summarising the properties of the four fields
comprising WTS. }
\label{tab:fields}
\end{table}

Both transit surveys \citep[eg.][]{Berta2012,Miller2008} and eclipsing
binary studies \citep[eg.][]{Morales2010} suffer from the intrinsic
photometric variability of the target stars which must be constrained
and accounted for. A thorough understanding of activity of stars is
vital in order to uniformly obtain the parameters of any orbiting
planets from both transit and radial velocity surveys, or indeed the
thresholds set by activity beyond which it is impossible to detect
planets. In this paper we classify a sample of M dwarfs in the WTS
survey and search for periodic variability in the long-baseline
$J$-band light curves. While we do not intend to directly test for
planet parameter retrieval, we aim to corroborate previous surveys of
M dwarf variability at shorter wavelengths in the near infra red and
parameterise the variability of the sample.

\section{M dwarf variability}\label{sec:M dwar variability} The
periodic variability in M dwarfs can result from the modulation of
brightness by cool star spots. M dwarfs, like all stars, enter the
main sequence rotating rapidly due to the angular momentum of the
cloud from which they form. The rotation induces a dynamo giving rise
to magnetic activity in the photosphere that produces stellar
spots. For solar-type stars the $\alpha - \Omega$ dynamo generated at
the tachocline, the boundary between the convection and radiative
zones, \citep{Browning2010} results in latitude dependent spots from
the generation of toroidal fields by differential rotation within the
star \citep{Brown2008} that confine magnetic flux tubes to high
latitudes \citep{Moreno1992,Schuessler1992}, and \citet{Granzer2000}
find that low latitude magnetic flux emergence can occur in very
young, partially radiative, main sequence stars, which could give rise
to spots at all latitudes. However, stars later than M3.5 are fully
convective, so no tachocline exists at which such a dynamo can be
generated, yet stars of a later spectral type are still active and
this activity may be dependent on an alternative form of dynamo
\citep{Rockenfeller2006,Brown2008}. This switch over at M3.5 has been
confirmed through observations of the magnetic energy of M dwarfs
\citep[e.g.][]{Reiners2008}. Indeed, spectropolarimetric studies of
objects known to be fully convective by \citet{Morin2008,Morin2010}
have found such a switch in magnetic field morphology. An alternative
dynamo, ultimately dependent on the Coriolis force and called
$\alpha^2$, has been shown by \citet{Chabrier2006} as being the type
of dynamo process occurring in fully convective, low mass objects such
as late M dwarfs and brown dwarfs. The magnetic field generated by the
$\alpha^2$ dynamo is quadrupolar. The more distributed field in these
models could lead to a more distributed spot pattern. The exact nature
of the dynamo process and its dependencies in fully convective M
dwarfs is yet to be properly understood, while observations continue
to place restraints on future models.

Observations by \citet{Barnes2003} indicate a uniform distribution of
star spots at all latitudes for partially radiative stars, while
others find spots concentrated in the polar regions
\citep{Morales2010,Jeffers2007} that more closely resemble model
predictions of magnetic activity. \citet{Frasca2009} fit K and M
dwarfs light curves of varying morphologies using a variety of models
consisting of just two large spots, although they do not consider
alternative spot distributions and instead vary the spot latitudes and
temperatures, and the stellar inclination. An approximately uniform
distribution of spots will also result in a variable light curve due
to local over- or under-densities of spots on the star. Photometric
observations provide limited information about precise starspot
morphology, but remain useful for characterising general trends in
active, rotating stars.
 
Rotation and activity have been shown to be correlated in M
dwarfs. Based on observations on a sample of 123 M dwarfs
\citet{Browning2010} find that all fast rotating stars are active
(while not all active stars are fast rotating). \citet{Messina2003}
study 274 cool, main sequence stars in clusters and find relations
between $V$ band amplitude, ratio of X-ray luminosity to bolometric
luminosity, period, and Rossby number (the ratio of the rotation
period to the convection turnover time). They find a period-amplitude
relation for late type dwarfs (K6 to M4) with two upper envelopes,
with a general trend of decreasing amplitude for larger periods, with
a similar relation found for amplitude and Rossby
number. \citet{Messina2011}, however, do confirm the link between
cluster age and the rotation period. They find 75 periodic variable
main sequence stars, including 30 late type variable dwarf stars
(earlier than K2, and therefore partially radiative), in the open
cluster M11, and find a mean period of 4.8 days, compared to 4.4 days
for the younger M35 cluster and 6.8 days for the older M37 cluster.
The periods found by \citet{Messina2011} range between 0.2214 and
21.05 days over an 18 day long observation period. Radial velocity
work by \citet{Reiners2007} confirms the activity rotation connection
up to the fully convective boundary.  \citet{Rockenfeller2006} find 19
variable field M dwarfs (M2 to M9) in the $I$, $R$ and $G$ bands. They
find stars of spectral types M2.5 to M9, and periods for some between
3.3 and 13.2 days. The fraction of variable stars among these field M
dwarfs is $0.211\pm0.11$ \citep[see references in][]{Rockenfeller2006}
and a trend of increasing light curve amplitude for later spectral
types is seen, with stars later than M10 having the greatest magnitude
variation. They find for the M9 star 2M1707+64 a spot coverage
fraction of less than 0.075, with a spot temperature to photospheric
temperature difference of 4\%-7\%, corresponding to 200K to 300K
cooler spots, and a light curve amplitude of ~0.04 in
magnitude. Frasca et al (2009) also fit somewhat greater values of
spot to photosphere temperature difference in their two-spot models of
variable M dwarfs, ranging from 9\% to 29\% cooler spots, and in one
instance, spots 10\% hotter than the photosphere, possibly may be due
to continued accretion on to the star.

\section{Variable Late-Type Dwarf Selection}\label{sec:Selection}
\subsection{Colour selection and spectral typing}\label{sec:Typing}

The spectral types of the majority of stars in the WTS data are
unknown. Although colour cuts have shown to be a useful tool for
extracting samples of M dwarfs while eliminating sources of
contamination \citep{Plavchan2008}, and estimation of spectral
subtypes was preferential for analysing relations between spectral
type and other parameters. We therefore made spectral classifications
using colour-spectral type relations to identify the WTS M dwarfs.
\citealt{Covey2007} find the location of main sequence stars across
the Morgan-Keenan spectral types from near infrared Two Micro All Sky
Survey (2MASS) $JHK_s$ and visual Sloan Digital Sky Survey (SDSS)
ugriz photometry. They find the median position of stars across
spectral type O to M for main sequence, giant and supergiant stars in
a 7 dimensional colour space and the relationship between spectral
type and colour. Using these relationships it is possible to estimate
both a spectral type and luminosity class for any given stars
providing they are observed in SDSS. Once the spectral types for all
the WTS stars had been estimated a sample of M dwarfs could be
extracted.

$JHK_s$ magnitudes for many of the stars are available from both the
2MASS and WTS surveys, the WTS magnitudes being the more accurate and
complete. In order to use the relations from \citet{Covey2007} the WTS
magnitudes were converted from the Mauna Kea Observatory (MKO) system
to the 2MASS system using empirical relations found by
\citet{Hewett2006}. The 03hr, 07hr and 19hr fields fully overlap with
SDSS observations, and $\sim96\%$ of the $J$$<$17 objects have
corresponding SDSS magnitudes, whereas the 17hr field is not covered
by SDSS, and a spectral type was only assigned to those with SDSS
observations.

On SDSS/2MASS $ugrizJHK_S$ colour-colour diagrams the loci of main
sequence and giant stars are discernible enough to fit discrete
tracks. We use the relative positions of each WTS star to these tracks
to estimate whether a star is on the the main sequence or not. Regions
where the tracks are discrete are found on plots of the colours of
consecutive passbands: $r-i$ versus $g-r$, $z-J$ versus $i-z$, $J-H$
versus $z-J$, $H-K_s$ versus $J-H$; additionally colours of wider
baselines were used: $i-K_s$ versus $g-i$, $r-z$ versus $u-r$, $r-z$
versus $J-H$. The use of multiple colour-colour plots allows for
greater robustness in the identification. Polynomial functions are
least-squares fitted to each of the trends to create tracks for giants
and dwarfs in colour-colour space. The orders of the polynomials is
determined such that it is the highest degree before there is
over-fitting, determined from the r.m.s. residual values. A quantified
comparison between a star's colours and the dwarf and giant tracks in
each diagram can then be made. A star's separation in colour space
from each track is found and divided by the 2$\sigma$ error on the
star's colour index to identify to which track a star lies
closest. Stars that lie within 2$\sigma$ of one of the polynomial fits
are assigned as being a dwarf or giant respectively, whereas all other
stars are flagged as being ambiguous under that colour-colour
regime. The modal identification from the comparisons with the seven
colour-colour relations is found giving each star an overall
categorisation as either dwarf, giant or ambiguous. As most stars
occupy a degenerate position on the tracks this serves primarily to
eliminate any star from the sample that can be positively identified
as being a giant.

\begin{figure} \centering
\includegraphics[angle=270,width=1.0\columnwidth]{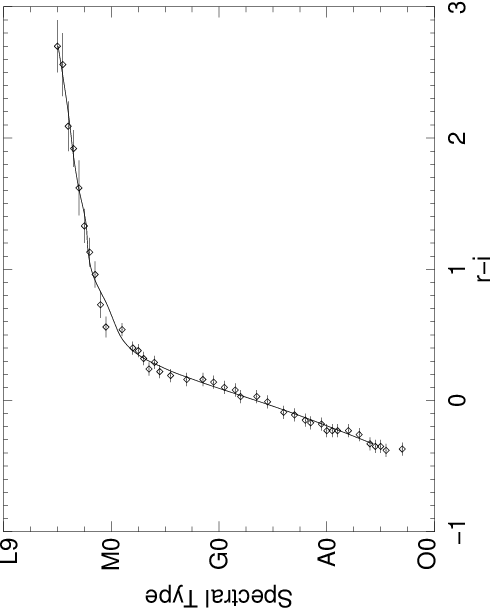}
\caption{An example plot of the colour-spectral type relations used to
estimate spectral types for the WTS sample. The points are the colour
indices for types, earlier than M0 from \citet{Covey2007} and for
later types from \citet{West2011} which extend the series to M10. The
fitted spline was used to infer spectral types for the stars in the
WTS sample. See text for a full list of colours used.}
\label{fig:ristcolspec}
\end{figure}

\begin{figure} \centering
\includegraphics[width=1.0\columnwidth]{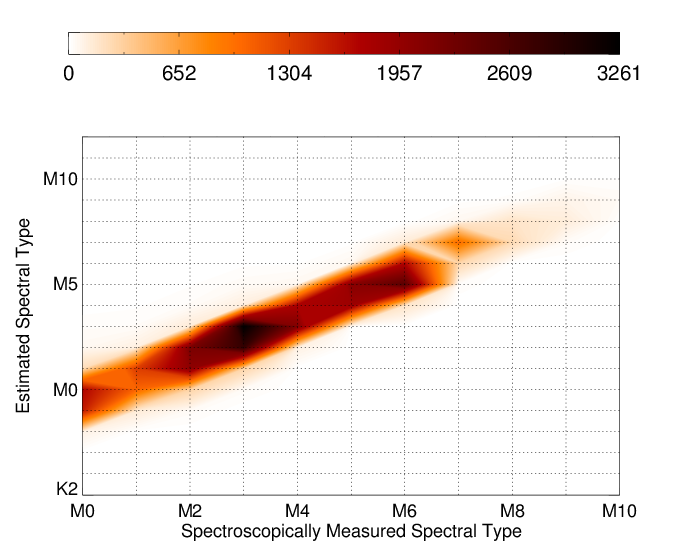}
\caption{A 2-D histogram of the spectral types of known M dwarfs
versus the spectral types found from the methods presented here,
showing the accuracy of our spectral typing is $\sim\pm1$. A colour
version of this figure is available in the online article.}
\label{fig:hist}
\end{figure}

In addition to the luminosity class, the spectral type can be
estimated from the colour indices using the colour-spectral type
relation presented by \citet{Covey2007}, and \citet{West2011} who
provide the relation extended to later spectral types. A method for
finding the spectral type is similar to that used to find the
luminosity class; smoothing splines were fitted to the colour
indices. The colour indices used were $r-i$, $i-z$, and $z-J$
(e.g. Figure \ref{fig:ristcolspec}); other SDSS and 2MASS colours
being either too metallicity dependent or having too large
uncertainties. The mean and standard deviation of the spectral types
for each colour index give an estimation for spectral type of each
star in the sample. To assess the reliability of our classification,
we applied the method to M dwarfs of known spectral types from
catalogues published by \citet{West2011}. Figure \ref{fig:hist} shows
that the match between our estimated spectral type and the
spectroscopically determined spectral type follows a 1:1 relation with
an error of approximately $\pm$1 sub-type.

One constituent field of the WTS, the 17hr field, was not observed in
the SDSS, and therefore spectral type estimates were not made,
although a variability selection described in the following section
was made on the reddest stars as defined by our initial colour cut at
$H-K > 0.175$ (corresponding approximately to K7V and later), combined
with colours cuts presented by \citet{Plavchan2008} to eliminate M
giants and background, reddened sources.

\begin{figure} \centering
\includegraphics[ width=1.0\columnwidth]{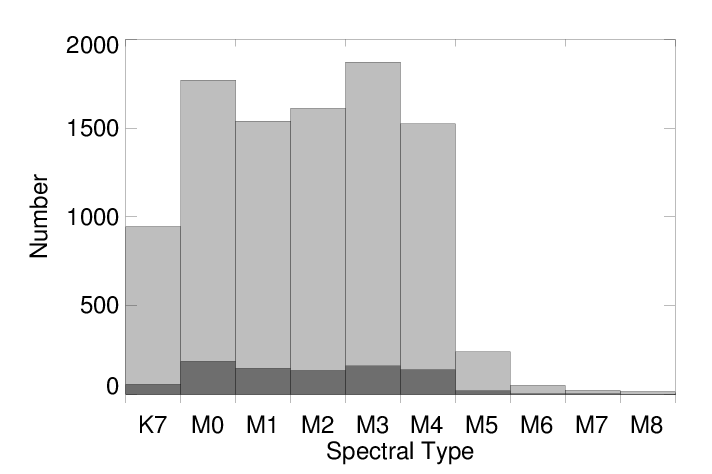}
\caption{Histograms showing the frequency of each spectral type in the
sample. The light grey is all stars with magnitude $J$$<$17, the dark
grey $J$$<$15.}
\label{fig:stfreq}
\end{figure}

\subsection{Variability Selection}\label{sec:Variability Selection}

\begin{figure*}
\begin{minipage}{18cm}
\includegraphics[scale=0.7, clip]{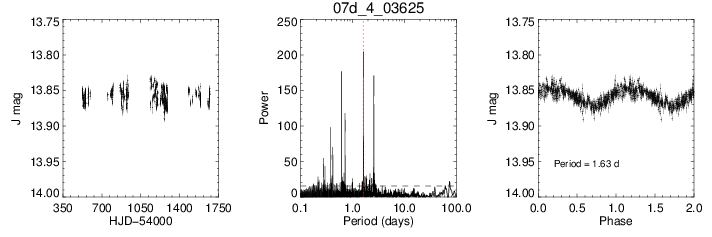}
\end{minipage}
\caption{Example light curve and periodogram for variable star: \it
From left to right: \rm Observed $J$ band time series for star;
Lomb-Scargle periodogram of time series between period window of 1 and
100 days with the maximum peak at 1.63 days. The dashed horizontal
line represents the 99\% confidence level, the red dashed line
represents the peak corresponding to the period about which the
lightcurve is to be folded; Phase folded light curve about period of
14.19 days with uncertainties omitted.}
\label{fig:example}
\end{figure*}

To identify periodically variable stars the frequency spectrum for
each of the M dwarf light curves in the sample is found using the
Lomb-Scargle periodogram, which is optimised for finding sinusoidal
shaped periodic signals \citep{Lomb1976,Scargle1982}. The periodogram
calculates a normalised power spectrum across a defined range of
periods; the peaks corresponding to periodic variations in the light
curve. It is robust with respect to a varying amplitude in that while
a linear change in amplitude will attenuate a signal and a periodic
change in amplitude will introduce a second signal, the signal
corresponding to the primary variability period will still be
detected. A false alarm probability (FAP, as described by
\citet{Horne1986}) threshold is determined to enable the peaks in the
periodogram to be identified with a given level of confidence and to
filter out peaks that are probably caused by noise. Our threshold is
defined such that all peaks in the periodogram with a value greater
than this value corresponds to a real signal with 99 per cent
confidence or better, that is to say, we discard by default any peaks
that are less probable than one generated by Gaussian noise by
determining the power of such a peak. All other peaks are ignored and
the remaining stars are retained as candidates for periodic
variability. This process only eliminates peaks that are unlikely to
be real signals, but leaves other signals that are not due to
modulation by photospheric spots. These other sources of periodic
variability are identified and eliminated from the sample. Potential
sources that are discernible are identified as: \newcounter{qcounter}
\begin{list}{\roman{qcounter}}{\usecounter{qcounter}}
\item Eclipsing binaries
\item Other variable stars
\item Periods imposed by the rotation of the Earth
\item Periods imposed by sampling rates
\item Periods from seeing variations with merged sources
\end{list}

One star was excluded as it was blended with the diffraction spike
from a known Mira type star V1134 Cyg \citep{Miller1966}, evoking a
period of 327.01 days. We discount all other potential identities for
the variables for having too early a spectral type or too long a
period as with RR Lyrae type stars or too large a light curve
amplitude that would be indicative of semi-regular, Mira or other
variable giant stars are excluded by the aforementioned colour cuts. A
total of 12 suspected or know eclipsing binaries were identified in
the sample as having light curves with periodic troughs of two
distinct depths, indicative of the primary and secondary eclipse of
the system, or periodic magnitude variations of $\Delta$$J$$>$0.02. We
find amplitudes larger than this necessitate unphysical photospheric
spot configurations using a light curve synthesis detailed in Section
\ref{sec:Spot morphology}. These were excluded from the sample.

Light curves with periods corresponding to $\sim$1 day are identified,
as is a sample of light curves with periods corresponding, within 3
significant figures, to fractions of a day (i.e. 1/2, 1/3, 1/4, 1/5),
thought to be caused by repeated sampling at a 1 day period. These
stars are not excluded from the search, but these periods are, such
that these peaks in the periodogram are ignored while other peaks are
identified. Additionally each variable star candidate is inspected by
eye in the field to reduce the potential of having any variations in
magnitude invoked by variability in other sources in a crowded
field. For many light curves, the periodograms have multiple peaks
above the FAP threshold, in which case the light curve is folded about
the primary or highest peak. From the 68 light curves in the sample a
peak to peak amplitude for each light curve is obtained by binning the
folded lightcurve, while the period is obtained directly from the
highest peak in the periodogram not to have been previously
discounted, about which the light curve is folded.

Each light curve is then broken down in three separate consecutive
epoch ranges each containing a third of the observations and each
range is again tested for variability. While the significance of the
peak in the periodogram varied for each epoch range, its presence
ensures that the periodic signal is maintained in the photometry of
the star for the duration of observations. This process also enables
the discover of some examples of evolving variability as discussed in
Section \ref{sec:Periodicity}.

\section{Results}

\begin{figure}
\hbox{\includegraphics[scale=0.7,width=1.0\columnwidth]{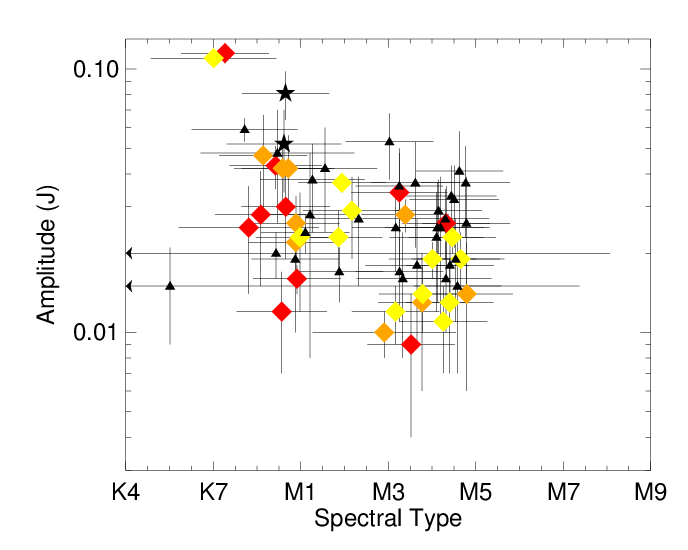}}
\caption{Estimated spectral type of each stars versus peak to peak J
band amplitude in the folded light curves. The uncertainties in
spectral type are the standard deviations of the spectral types found
per colour index. Errors in amplitude are estimated from the errors in
magnitude in the observed magnitudes. Symbols are described in the
text. The Spearman rank correlation coefficient (R$_S$=-0.35)
indicates a weak trend for decreasing amplitudes at later spectral
types. A colour version of this figure is available in the online
article.}
\label{fig:stamp}
\end{figure}

\begin{figure}
\hbox{\includegraphics[scale=0.7,width=1.0\columnwidth]{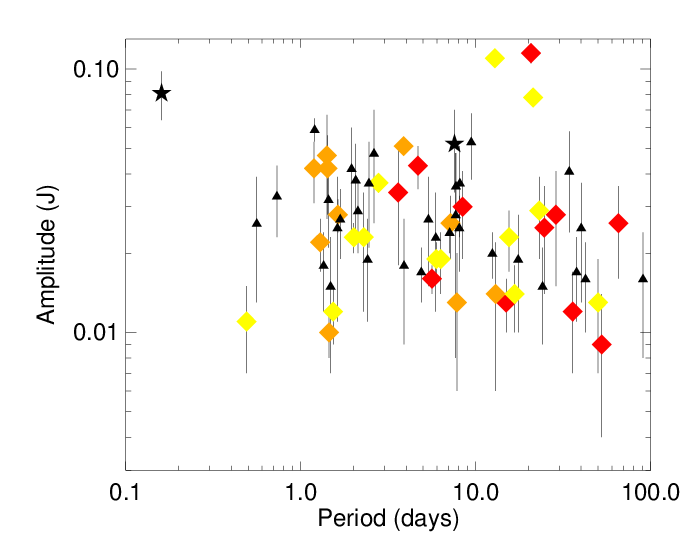}}
\caption{Variability period versus peak to peak amplitude. The periods
correspond to those in Figures \ref{fig:lc1} onward, amplitude as in
\ref{fig:stamp}. Symbols are described in the text. The Spearman rank
correlation coefficient (R$_S$=-0.24) indicates a weak trend for
decreasing amplitudes at longer periods; it is less correlated than
the trend found by \citet{Hartman2011}. A colour version of this
figure is available in the online article.}
\label{fig:pamp}
\end{figure}

\begin{figure}
\hbox{\includegraphics[scale=0.7,width=1.0\columnwidth]{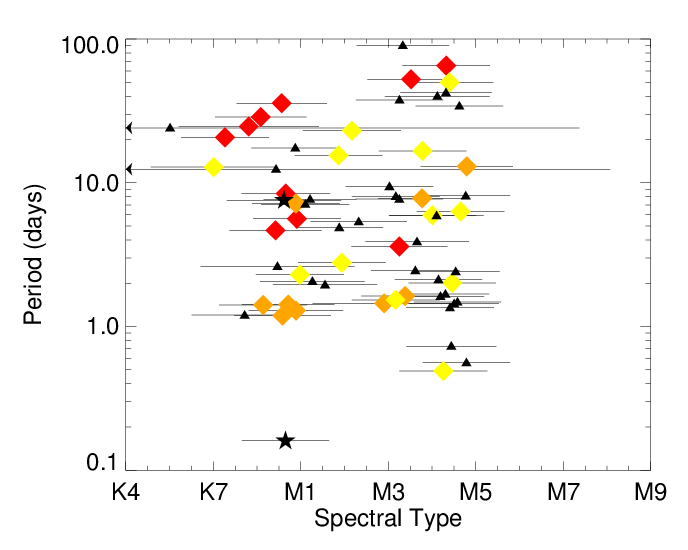}}
\caption{Estimated spectral type versus variability period. The
periods correspond to those in Figures \ref{fig:lc1} onward, spectral
type as in \ref{fig:stamp}. Symbols are described in the text. A
colour version of this figure is available in the online article.}
\label{fig:stp}
\end{figure}

\begin{figure*}
\begin{minipage}{18cm}
\hbox{\includegraphics[scale=0.7]{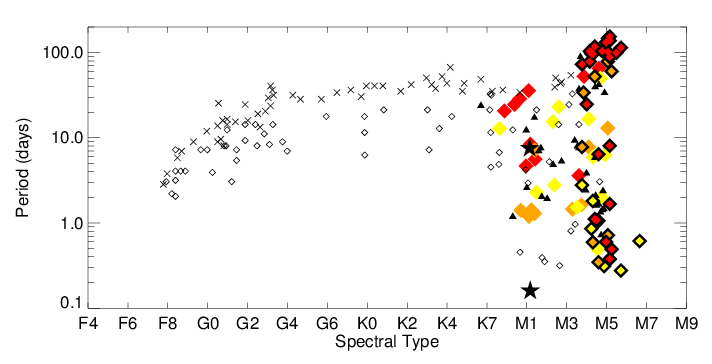}}
\end{minipage}
\caption{Estimated spectral type versus variability period, including
objects from the later-type sample, and compilation of rotating field
stars from the literature, presented in \citet[Figure 12]{Irwin2011}
for comparison. Symbols are described in the text, or else, the
symbols with black borders being those from \citet{Irwin2011} with
red, orange and yellow corresponding to their definitions of thick,
middle and thin disc populations respectively; crosses represent young
field stars ($\sim$1-2Gyr) and unfilled black diamonds represent old
field stars ($\sim$8-10Gyr). Spectral types were interpolated from the
standard spectral types and stellar masses \citep{Mamajek2011}. The
large v$_{tan}$ stars, which may be of the older, thick-disc
population, with spectral type M1V ($\sim$0.6 $M_{\odot}$) are shown
to exist with periods of $<$10 days, existing as outliers to the
sequence of old rotators. A colour version of this figure is available
in the online article.}
\label{fig:stpi}
\end{figure*}

The variability search yields 68 detections of variable stars (Table
\ref{tab:var}, Figures \ref{fig:lc1}-\ref{fig:lc3}) with colours
indicating they are late type dwarfs. A Spearman rank correlation
coefficient calculated for the amplitudes with respect to spectral
type finds a weak trend for lesser amplitudes at later spectral types
(R$_S$=-0.35, 99.7\% significance), with the largest amplitudes
detected from the earliest type stars. Figure \ref{fig:pamp} shows
amplitude against period, and while no upper envelope
\citep[e.g.][]{Messina2003} is found, a weak trend for smaller
amplitudes at longer periods is found using the same Spearman rank
coefficient (R$_S$=-0.24, 95.5\% significance), as \citet{Hartman2011}
find in sample of periodically variable M dwarfs in the HATNet survey,
although with a less defined and significant correlation. This would
be expected to arise from the more rapid rotation of the short period
variables driving an increase in magnetic activity, given the
dependence of the $\alpha-\Omega$ dynamo on rotation. It should be
noted that the random orientation of the axes of rotation of these
stars would attenuate this correlation by mitigating the difference in
flux observed along the line of site as an unevenly spotted star
rotates. For stars with randomly distributed inclinations we would
expect the peak amplitude to be on average $1/\sqrt{2}$ of the actual
amplitude. The switch between radiative to fully convective occurs at
types later than M3.5 \citep{Rockenfeller2006}, although this is not
visible in Figures \ref{fig:stamp}, \ref{fig:stp}, and \ref{fig:pamp}
due to the lack of variable stars later than M4 detected in our
sample.

We can take advantage of these results to make comparisons with
rotation-evolution studies, using tangential velocities ($v_{t}$)
obtained from SDSS proper motions and distance estimates from
photometric parallaxes \citep{Bochanski2011} as approximate tracers of
population \citep[eg.][]{Reiners2008}. We label stars with a
$v_{t}\gtrsim30$ kms$^{-1}$ as likely old disk stars (red diamonds in
Figures \ref{fig:stamp}, \ref{fig:pamp}, \ref{fig:stp},
\ref{fig:stpi}), and those with a $v_{t}<15$ as young disk stars
(yellow diamonds). Other stars are marked with orange diamonds. Two of
the faintest stars in our sample are found to have $v_{t}$=309$\pm$35
kms$^{-1}$ and $v_{t}$=814$\pm$79 kms$^{-1}$ (marked with black
stars), consistent with being members of the halo. We find our old
disk stars are primarily slower rotators, corroborating the results
and collated data of \citet{Kiraga2007}, and the later MV stars in
\citet{Irwin2011}, who find a greater number of slow rotators amongst
the old disk stars, and also find young disk stars occupying the full
range of detected periods. We find the majority our 'old' stars at
periods $>$20 days, in agreement with their conclusion of a more rapid
spin down time for more massive, partially radiative stars. Although
we do find 5 early MV stars with periods $<$10 days, this may be as a
result of the crude method of assuming an age from $v_{tan}$ that does
not take into account the full space velocity of the star, and these
stars may in fact be members of the younger population, as early M
types only remain active for $<$2Gyr \citep{West2008}.

\section{Discussion}\label{sec:Discussion}

\begin{figure*}
\begin{minipage}{18cm}
\hbox{\includegraphics[scale=1.0,width=1.0\columnwidth]{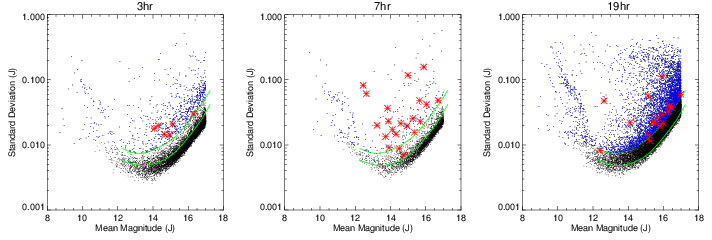}}
\end{minipage}
\caption{Dispersion (standard deviation) of each lightcurve versus
magnitude. Red symbols indicate the mean brightness of the star versus
the amplitudes in $J$ magnitude. The blue points are those that lay
above the green line indicating the effective 1$\sigma$ and 2$\sigma$
boundaries. The majority stars with the larger standard deviation show
detectable periodic variability but may be stochastically variable,
while others are eclipsing binaries. A colour version of this figure
is available in the online article.}
\label{fig:jvnoise}
\end{figure*}

\subsection{Completeness}\label{sec:Completeness} In our sample, the
lack of detected variable late-MV stars may be reflective of the very
low numbers of stars later then M4 (see Figure \ref{fig:stfreq})
present in the total sample, such that this lack of active stars
beyond M7 may purely be indicative of the bias in the sample towards
earlier spectral types, as we find no variables later than
M4V. \citealt{Ciardi2011} find a variability fraction of 0.367 for M
dwarfs over the same observation window based on the dispersion of
points in the lightcurves about the mean, but note that their sample
includes stars that are not periodic in their variability. They also
find that the time scales for such variability is on a time scales of
weeks or more. \citet{Rockenfeller2006} find the fraction of variable
stars among field M dwarfs to be $0.21 \pm 0.11$. For periodically
variable stars, we find a the variable fraction to be $\ge$0.007. For
comparison, \citet{Hartman2011} find a fraction of 0.057 stars as
reliable variables but 0.001 periodic and quasi-period variables
amongst field M and K dwarfs in the HATNet survey.

standard deviation of each 
magnitude. Considering only the stars above the one 
an average fraction per magnitude bin (12$<$J$<$17) 

In order to further assess the sensitivity of our period detection
method, and to probe how noise in the light curves affects detections,
we perform Monte Carlo simulations. These simulations are carried out
in such a way to also determine the effects of aliasing of the
observation periods with real periodic signals. Light curves with no
intrinsic periodic signal are randomised with respect to magnitude,
maintaining observation times. Sinusoidal signals of random periods
and amplitudes are then injected. The same Lomb-Scargle periodogram is
applied to each lightcurve and tested for a peak more significant than
the 0.1 FAP threshold, corresponding to the injected period, which by
our method would constitute a detection. The resulting simulations
show that the for 12$<$$J$$<$15 periods of less than 20 days with a
$J$ band variability greater than 0.01 should always be
detected. Indeed this constitutes the majority of periodic variability
detections despite a greater sample of stars at $J$$>$15. These
simulations indicate that that for 12$<$$J$$<$14 the fraction of
variable stars is is 2\%. This variable fraction is still are an order
of magnitude less than those found at surveys of shorter wavelengths,
and reasons for this are discussed in section \ref{sec:Spot
morphology}.

\subsection{Variability}\label{sec:Periodicity}

Periodicity only occurs for those M dwarfs that are active, and this
activity also results in flaring. \citet{Reiners2007} predicts that
rotational periods of the order of several weeks will be found for
early M dwarfs with difficulty, due to the relatively short duration
flaring events resulting in noise disguising the periodicity. The WTS
observations are however conducted in the $J$ band, which alleviates
this issue as flaring events are rarely as detectable in the near
infrared \citep{Rockenfeller2006}. \citet{Tofflemire2012} find no
evidence of flaring in $JHK_{S}$ pass bands during simultaneous
optical and infrared monitoring of three M dwarf stars for 47 hours
corresponding to the flaring events they detect in the $U$-band. We
have serendipitously found one particularly large flaring event from
the M4V($\pm 1$) star 19d\_1\_12692, of $\geq$$\Delta0.2$ in $J$ band
magnitude. The incomplete observations of the flare lasted for 49
minutes and thus the overall duration could have been several
hours. This brightness of the flare is in contrast to the results of
\citet{Tofflemire2012} who expect to find flaring events in the $J$
band occurring on mmag scales, and would transform to a $u$-band
response of $>$$\Delta6$, which by extrapolation of the frequencies of
flaring events per magnitude found by \citet{Davenport2012} would be
expected to be observed less than once a year. Due to such a low rate
of occurrence it is unlikely that such a flare would obfuscate a
planetary transit, but such high intensity flares may have
astrobiological significance.

\begin{figure}
\hbox{\includegraphics[scale=0.7,width=1.0\columnwidth]{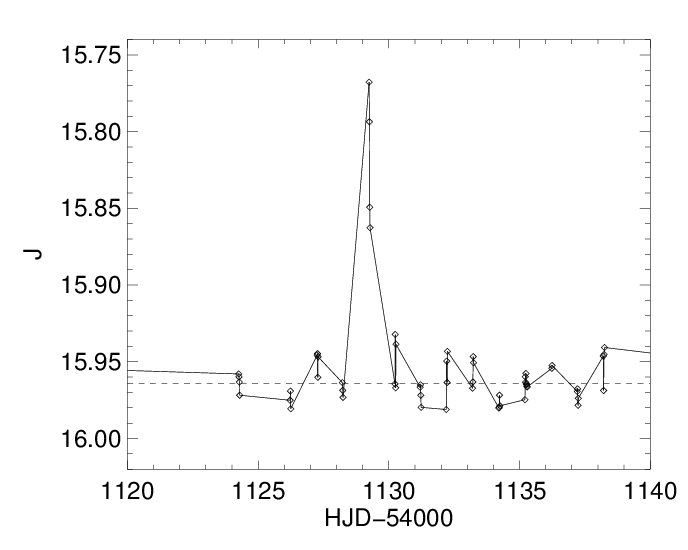}}
\caption{The flare observed from the variable M4V star
19d\_1\_12693. Consecutive observations are joined. The observation of
the flare is partial but those observations composing it were made
over a 49 minute period.}
\label{fig:flare}
\end{figure}

In our sample we have detected M dwarfs with periods between hours and
several weeks, with the largest period found at 50.02 days. For
comparison, \citet{Messina2011} find 30 late-type periodic variable
stars in the open cluster M11 with periods typically ranging between
$\sim$1 day to $\sim$20 days, whilst \citet{Messina2003} finds dwarf
stars with spectral types K4-M6 with periods between $\sim$1 and
$\sim$10 days. \citet{Irwin2011} similarly find long periods, in a
wavelength region approximately corresponding to the $i+z$ bands for M
dwarfs ranging between 0.28 and 154 days in their study of rotating M
dwarfs in the MEarth Transit Survey. \citet{Benedict1998} find
evidence for periodicity in Barnard's Star of $\sim$130 days and
Proxima Centauri of $\sim$40 and $\sim$83 days, the latter confirmed
by \citet{Kiraga2007}. Whilst the WTS periodic variable M dwarf sample
produced no convincing evidence of periodicity on these time scales, a
large number of periods $>$ 80 days were found in the periodicity
search and subsequently rejected when no sinusoidal variability was
seen. Indeed it is not expected that all spot coverage on a star will
induce a periodic variability in brightness if the spot coverage is
not sufficiently inhomogeneous to invoke a detectable difference in
brightness as the star rotates, or if the spot coverage changes in
morphology at shorter time scales than the rotation period, although
even in such a case a rotation period should be found by a
Lomb-Scargle periodogram. This is discussed further in the following
section.

\begin{figure*}\hbox{\includegraphics[scale=0.7,
clip]{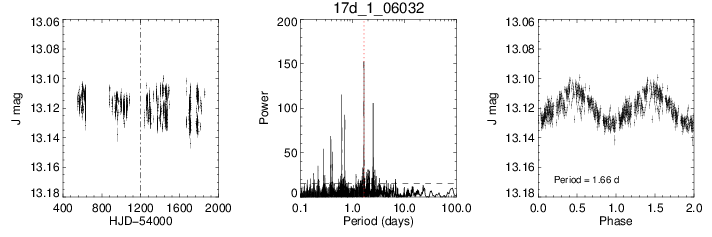}}
  \caption{As in Figure \ref{fig:lc1}. For 17d\_1\_06032, periodic
variability was only found from the latter 50\% of the observations,
as marked, while no significant peak corresponding to this period was
found when the periodogram was performed over all observations. Indeed
a change in the amplitude of the unfolded lightcurve is visible across
the marked division. This could be due to evolution in the spots on
the photosphere such that, during the course of the observations, the
spot coverage became less uniform so as to become variably bright.}
  \label{fig:17hr}
\end{figure*}

From our simulations discussed in Section \ref{sec:Completeness} and
shown in Figure \ref{fig:mcplots} there is substantial loss of
completeness for periods $>$20 days due to aliasing effects. In
observations with Kepler \citet{Ciardi2011} find that M dwarfs are
primarily variable over these time scales or longer. We are therefore
limited to detecting only a smaller subsection of M dwarfs before our
detection rate becomes sensitive to aliasing with observational
periods, despite the overall long base line of observation time.

\subsection{Spot Morphology}\label{sec:Spot morphology}

\begin{center}
  \begin{figure*} \centering
    \includegraphics[scale=0.6]{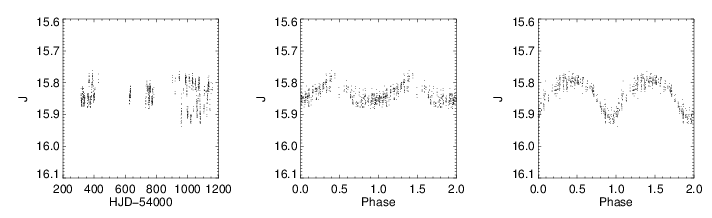}
    \includegraphics[scale=0.6]{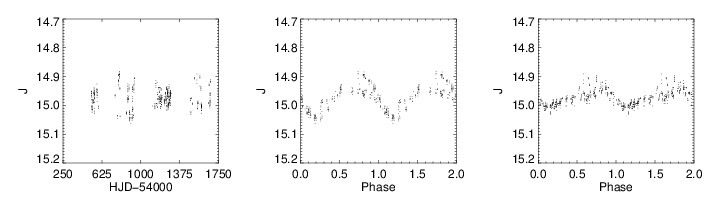}
    \caption{The upper plots show light curves with errors omitted of
the stars 19c\_2\_05428 and folded about the same period of 20.68
days, with the centre panel showing observations at HJD$<$54900 and
the panel on the left with observations at HJD$>$54900. The low plots
show and 07e\_2\_02466 and folded about the same period of 12.86 days,
with the centre panel showing observations at HJD$<$55000 and the
panel on the left with observations at HJD$>$55000. For 19c\_2\_05428,
the phase is conserved, although the two distinct amplitudes and
morphologies show the dichotomous nature of the variability over
time. This may be an indicator of longitude dependent spots or spot
groups evolving over time. 07e\_2\_02466 however exhibits a similar
saw tooth like morphology in both sets of observations but the
amplitude appears to vary more stochastically, although over hundreds
of days.}  \centering
    \label{fig:19cvar}
  \end{figure*}
\end{center}

We find some evidence for a change in spot coverage was found in one
star with an indeterminate spectral type in the 17hr field,
17d\_1\_06032 (Figure \ref{fig:17hr}), and in the stars 07e\_2\_02466
and 19c\_2\_05428 (Figure \ref{fig:19cvar}). In the former, no
significant peak in the periodogram, performed over the entire series
of observations, corresponded to an obvious periodicity in folded
light curves, thought the most significant peak in the periodogram for
only the later 50\% of the observations did correspond to a period for
which a folded light curve displayed obvious periodic behavior. In the
latter two a strong peak corresponding to the period is found in their
periodograms although the amplitude of the variation changes over
timescale of months, perhaps similarly indicative in a change in the
spottedness of the stars. The changes in both instances happened over
a time scale of $<$100 days, although due to a lack of observations
during the transition it is unknown what the true speed of the
amplitude change was. In both 17d\_1\_06032 and 19c\_2\_05428 however,
the change in amplitude was dichotomous and does not appear to vary
much once the new amplitude if established, whereas in 07e\_2\_02466
the amplitude varies over hundreds of days.

Since little information is known about the distribution of spots on
any of the M dwarfs in our sample, a spot model is used to generate a
range of scenarios against which the results presented here can be
compared. The simulations were generated by the Doppler Tomography of
Stars code (DoTS) \citep{Cameron1997} that can produce synthetic light
curves at a chosen wavelength (13000\r{A}, corresponding approximately
to the $J$ band midpoint) for a range of photospheric and spot
temperatures ($T_p$, $T_s$ respectively) and spot coverage models. The
spot coverage models are identical to those used by
\citet{Barnes2011}, including seven models in total covering solar
minimal and maximum coverage to more extreme cases for active stars
\cite[see Table 1,][]{Barnes2011}. Three $T_p/T_s$ ratios are used for
each coverage model, a solar like ratio of 0.6, a less extreme ratio
of 0.8, and an adjustment such that $T_s = T_p - 250$ K, comparable to
models fitted to variable field M dwarfs in $G$, $R$, $I$ bands by
\citep{Rockenfeller2006}. In addition to these models a single spot
model at the 0.6 contrast ratio and -250K contrast were simulated for
comparison with randomly distributed spot models. In these models a
single 5$\degree$ wide spot was places at 0$\degree$ latitude on a
star of zero inclination. Each simulation is carried out for a range
of effective temperatures corresponding from spectral types M0V to
M10V at 250K intervals, based on mass-spectral class models by
\citet{Baraffe1996} and a spectral type-effective temperature scale
derived by \citet{Mohanty2003}.

From the simulations, synthetic light curves are obtained giving a
range of peak to peak amplitudes for each scenario of spot coverage,
against which the variable stars in the sample can be compared in
order to place limits on the spot coverage of each star. The results
of the DoTS simulations are plotted in Figure \ref{fig:dots}. These
simulations place some limits on the spot coverage over the variable M
dwarfs found here that suggest a star spot coverage modeled at 18\%
and 48\% for lower $T_s/T_p$ ratios and a lower coverage model at
6.1\% for the larger $T_s/T_p = 0.6$ model. Also included in this
range are many of the single spot models, although the morphology of
these is distinct from a sinusoid and we can rule out such models as
an explanation for the variability we detect. There exists numerous
degeneracies between models and scenarios not modeled here rendering
further determinations of spot coverage difficult. The single spot
model included a spot of 5$\degree$ width centered at 0$\degree$
latitude, although larger spots at changing latitudes could generate
the same peak to peak amplitude, and there maybe be multiple spots or
spotted regions. The light curves for these configurations would
differ in shape, as modeled by \citet{Frasca2009}, but such shape is
difficult to distinguish due to noise in the WTS light curves, in
addition with the potential for individual spots to change in size
over the duration of the survey. Furthermore, the amplitudes are
highly dependent on the $T_s/T_p$ ratio which introduces extra
degeneracy. While a well defined spot coverage cannot be determined,
it is apparent that many of the lightcurves are not sinusoids, instead
resembling a saw tooth morphology
(Figures~\ref{fig:lc1},~\ref{fig:lc2},~\ref{fig:lc3}), suggesting a
spot density that is more sharply dependent on longitude.

The detection of stars with amplitudes $\Delta$J$>0.1$ is at least
evidence for spot coverage $\gtrsim$50$\%$, as required to explain
observations of M dwarf radii by \citet{Jackson2009}, assuming a more
random distribution of spot coverage, or large variation in the
spottedness of such a star with respect to longitude.Additionally,
some information can be extracted from comparisons between the $V$,
$I$ and $J$ band amplitudes. The amplitudes in the $V$ and $I$ bands
are greater for stars with lower spot coverages which may explain
discrepancies between the variable fraction found here and the higher
fractions found in the aforementioned surveys. Simulations show that
$J$ band amplitudes for any given spot induced variability are 0.44 of
those in the $V$ band and 0.55 of those in the $I$ band. Furthermore,
the simulations indicate we are limited to detecting only stars with
spot coverages greater than $\sim$10 per cent, as variability possible
under lesser spotted models is of a lesser amplitude then the noise
limit of WTS, and affirms the use of the near infra-red for planetary
transit searches as spot induced variability is of a lesser magnitude
that that observed in shorter wavelengths. Indeed, in variable samples
observed at shorter wavelengths such as \citet{Rockenfeller2006} ($R$,
$I$) and \citet{Irwin2011} ($\sim$$i+z$ bands) find lightcurve
amplitudes of the order of $\sim$0.005-0.01 in their respective pass
bands. In the sample described in \citet{Rockenfeller2006} only 16\%
of the detected $I$ band amplitudes are $>$0.02 in magnitude. Lower
level variability scaled to the $J$ band would be less than the range
detectable in our sample due to our noise limitations found in our
Monte Carlo simulations, which reflect the general noise limit of the
WTS (Figure \ref{fig:jvnoise}). This, coupled with aliasing
sensitivity for periods $>$20 days, may explain the order of magnitude
lower fraction of periodically variable M dwarfs detected in our $J$
band survey.

\section{Conclusions}\label{sec:Conclusions}

In this paper we have selected a sample of M dwarfs from the WFCAM
Transit Survey by estimating their spectral type using colour
indexes. From this a sub-sample of variable objects are selected using
a Lomb-Scargle periodogram. 68 variable rotating MV stars are found
with a range of periods from 0.16 to 90.03 days and amplitudes from
0.009 to 0.115 in the $J$ band. Additionally three others with a less
constrained spectral type are found. We infer population membership
from tangential velocities and find our results to be in agreement
with previous such studies.

Using a lightcurve synthesis code we find that these stars may have a
high degree of spottedness, of the order of 10 per cent surface
coverage or more, in some cases $\gtrsim$50 per cent, at least at time
scales of $<20$ days, and we estimated a fraction for these variable M
dwarfs of at least of 1 per cent from the most complete
subsamples. These results indicate that transit surveys carried out in
the $J$ band may be less susceptible to the effect of spot induced
variability in photometric observations, as our simulations suggest
they are predisposed to only detecting the variability of the most
active of M dwarfs. This would minimise intrinsic variability and
allow the better parameterisation of transiting planet and eclipsing
binary systems by observing in the near infrared. One example of a
particular large flare in the $J$ band is serendipitously found. We
also find evidence for evolving spot morphologies in the form of light
curve amplitudes varying over periods of months in three of stars. In
one late type dwarf of unknown spectral type periodic variability was
found to switch on after months of inactivity.

\section{Acknowledgments}\label{sec:Acknowledgements} GB and BS are
supported by Rocky Planets Around Cool Stars (RoPACS), a Marie Curie
Initial Training Network funded by the European Commission's Seventh
Framework Programme.  \\\\ NG, JB, DP, SH, JB, HJ, CdB, M-CG-O, SN and
EM have received support from RoPACS during this research, a Marie
Curie Initial Training Network funded by the European Commission's
Seventh Framework Programme.  \\\\ This work was partly funded by the 
by the Funda\c{c}\~{a}o para a Ci\^{e}ncia e a Tecnologia (FCT)-Portugal 
through the project PEst-OE/EEI/UI0066/2011 \\\\ We would like to thank
collaborators across the RoPACS network for their useful comments. We
acknowledge a useful contribution from Peter Hensmen towards this
project.


\appendix
\section{} 
\begin{table*}\scriptsize
\begin{tabular}{ | l | l | l | l | l | l | l | c | c | c | } \hline
Name & RA$^a$ & Dec$^a$ & J$^b$ & ST Est.$^c$ & P$^d$ & a$^e$ &
PM$_\alpha^f$ & PM$_\delta^f$ & v$_t^g$ \\ \hline
03d\textunderscore1\textunderscore04395 & 55.115972 & 39.134701 &
14.32 & M1 $\pm$ 1.01 & 4.880 & 0.017 $\pm$ 0.004 & --- & --- & --- \\
03e\textunderscore1\textunderscore02053 & 53.784274 & 38.954377 &
14.95 & K5 $\pm$ 9.36 & 24.100 & 0.015 $\pm$ 0.006 & --- & --- & ---
\\ 03e\textunderscore2\textunderscore03325 & 54.204782 & 38.846530 &
14.08 & M1 $\pm$ 1.01 & 7.090 & 0.024 $\pm$ 0.004 & --- & --- & --- \\
03f\textunderscore1\textunderscore03040 & 54.006089 & 38.807408 &
16.30 & M4 $\pm$ 1.01 & 8.110 & 0.037 $\pm$ 0.014 & --- & --- & --- \\
03h\textunderscore2\textunderscore01177 & 55.772275 & 38.959886 &
14.64 & K9 $\pm$ 7.64 & 12.440 & 0.020 $\pm$ 0.004 & --- & --- & ---
\\ 03h\textunderscore4\textunderscore04646 & 55.184813 & 39.423991 &
15.13 & M3 $\pm$ 1.00 & 37.690 & 0.017 $\pm$ 0.006 & --- & --- & ---
\\ 07b\textunderscore2\textunderscore02081 & 106.059460 & 12.868579 &
15.69 & M4 $\pm$ 1.00 & 2.120 & 0.029 $\pm$ 0.009 & --- & --- & --- \\
07b\textunderscore2\textunderscore02125 & 106.158970 & 12.867247 &
14.10 & M3 $\pm$ 1.00 & 16.660 & 0.014 $\pm$ 0.004 & -1.410 $\pm$
2.642 & -1.606 $\pm$ 2.642 & 3.047 $\pm$ 0.624 \\
07b\textunderscore2\textunderscore02237 & 106.272010 & 12.863213 &
13.24 & M4 $\pm$ 1.00 & 5.950 & 0.019 $\pm$ 0.003 & -16.955 $\pm$
2.547 & 6.120 $\pm$ 2.547 & 15.996 $\pm$ 1.925 \\
07b\textunderscore3\textunderscore00631 & 106.072930 & 13.345031 &
14.31 & M4 $\pm$ 1.01 & 0.490 & 0.011 $\pm$ 0.004 & -1.657 $\pm$ 2.853
& -1.733 $\pm$ 2.853 & 3.317 $\pm$ 0.705 \\
07b\textunderscore3\textunderscore02281 & 106.137370 & 13.257735 &
15.64 & K9 $\pm$ 1.06 & 4.670 & 0.043 $\pm$ 0.008 & -4.709 $\pm$ 2.797
& -3.696 $\pm$ 2.797 & 49.275 $\pm$ 9.840 \\
07c\textunderscore2\textunderscore00282 & 106.827750 & 12.938019 &
13.71 & G9 $\pm$ 2.70 & 14.920 & 0.013 $\pm$ 0.003 & 0.142 $\pm$ 2.421
& -3.279 $\pm$ 2.421 & 115.837 $\pm$ 22.843 \\
07c\textunderscore2\textunderscore00294 & 106.899370 & 12.937313 &
13.91 & M3 $\pm$ 1.00 & 1.530 & 0.012 $\pm$ 0.003 & 9.569 $\pm$ 2.617
& -2.292 $\pm$ 2.617 & 16.275 $\pm$ 2.652 \\
07c\textunderscore4\textunderscore04250 & 106.459880 & 13.336219 &
14.76 & M3 $\pm$ 1.00 & 52.490 & 0.009 $\pm$ 0.005 & 1.724 $\pm$ 2.924
& -19.953 $\pm$ 2.924 & 42.274 $\pm$ 9.217 \\
07d\textunderscore1\textunderscore02956 & 106.627350 & 12.859053 &
13.83 & M1 $\pm$ 1.00 & 2.780 & 0.037 $\pm$ 0.003 & -3.240 $\pm$ 2.481
& -1.363 $\pm$ 2.481 & 7.927 $\pm$ 1.486 \\
07d\textunderscore1\textunderscore03180 & 106.617910 & 12.780124 &
16.04 & M0 $\pm$ 1.11 & 1.190 & 0.042 $\pm$ 0.011 & -3.872 $\pm$ 2.970
& -0.899 $\pm$ 2.970 & 37.971 $\pm$ 8.498 \\
07d\textunderscore1\textunderscore06063 & 106.510690 & 12.913766 &
15.35 & M3 $\pm$ 1.00 & 7.800 & 0.013 $\pm$ 0.007 & 8.695 $\pm$ 3.114
& -1.453 $\pm$ 3.114 & 22.054 $\pm$ 4.435 \\
07d\textunderscore4\textunderscore03625 & 106.725960 & 13.303723 &
13.87 & M3 $\pm$ 1.00 & 1.630 & 0.028 $\pm$ 0.004 & 5.903 $\pm$ 2.561
& -16.933 $\pm$ 2.561 & 27.624 $\pm$ 4.870 \\
07d\textunderscore4\textunderscore04577 & 106.692980 & 13.339136 &
16.70 & K9 $\pm$ 1.01 & 1.410 & 0.047 $\pm$ 0.020 & -1.171 $\pm$ 3.130
& 1.261 $\pm$ 3.130 & 25.460 $\pm$ 6.040 \\
07e\textunderscore2\textunderscore01022 & 105.870700 & 12.686058 &
12.45 & K2 $\pm$ 2.20 & 21.270 & 0.078 $\pm$ 0.002 & -3.729 $\pm$
2.322 & 1.183 $\pm$ 2.322 & 12.746 $\pm$ 2.243 \\
07e\textunderscore2\textunderscore02466 & 105.869530 & 12.632173 &
14.99 & K7 $\pm$ 1.44 & 12.860 & 0.110 $\pm$ 0.007 & -1.946 $\pm$
2.392 & -0.524 $\pm$ 2.392 & 16.488 $\pm$ 3.097 \\
07e\textunderscore2\textunderscore06232 & 105.994080 & 12.512096 &
15.27 & M0 $\pm$ 1.01 & 2.290 & 0.023 $\pm$ 0.011 & 0.327 $\pm$ 2.764
& -0.321 $\pm$ 2.764 & 2.618 $\pm$ 0.580 \\
07f\textunderscore3\textunderscore00675 & 106.076750 & 13.042346 &
12.62 & G5 $\pm$ 20.00 & 3.870 & 0.051 $\pm$ 0.002 & -11.750 $\pm$
2.322 & -1.021 $\pm$ 2.322 & 34.156 $\pm$ 4.526 \\
07f\textunderscore3\textunderscore03235 & 106.172640 & 13.106071 &
14.94 & M4 $\pm$ 1.00 & 6.310 & 0.019 $\pm$ 0.005 & -4.006 $\pm$ 3.053
& -6.566 $\pm$ 3.053 & 13.356 $\pm$ 2.988 \\
07g\textunderscore1\textunderscore04615 & 106.386640 & 12.613818 &
14.49 & M0 $\pm$ 1.03 & 35.790 & 0.012 $\pm$ 0.005 & -14.678 $\pm$
2.514 & -8.399 $\pm$ 2.514 & 77.421 $\pm$ 10.105 \\
07h\textunderscore1\textunderscore03267 & 106.617800 & 12.550436 &
14.59 & M0 $\pm$ 1.08 & 1.290 & 0.022 $\pm$ 0.005 & 0.743 $\pm$ 2.616
& -5.523 $\pm$ 2.616 & 22.316 $\pm$ 4.654 \\
19a\textunderscore1\textunderscore00838 & 292.765130 & 36.419518 &
15.99 & K8 $\pm$ 1.61 & 24.650 & 0.025 $\pm$ 0.011 & 1.516 $\pm$ 2.956
& -3.306 $\pm$ 2.956 & 39.212 $\pm$ 3.310 \\
19a\textunderscore1\textunderscore10932 & 292.511620 & 36.308370 &
16.08 & M0 $\pm$ 1.02 & 8.410 & 0.030 $\pm$ 0.011 & 5.269 $\pm$ 4.318
& 2.332 $\pm$ 4.318 & 52.077 $\pm$ 6.235 \\
19a\textunderscore3\textunderscore01981 & 293.095110 & 36.809146 &
15.64 & M4 $\pm$ 1.00 & 1.680 & 0.027 $\pm$ 0.008 & --- & --- & --- \\
19a\textunderscore3\textunderscore10735 & 293.304660 & 36.915622 &
15.14 & K8 $\pm$ 1.21 & 1.200 & 0.059 $\pm$ 0.006 & --- & --- & --- \\
19a\textunderscore3\textunderscore11735 & 293.329960 & 36.744529 &
16.78 & M0 $\pm$ 1.00 & 0.160 & 0.081 $\pm$ 0.017 & -36.959 $\pm$
4.379 & 51.498 $\pm$ 4.379 & 813.932 $\pm$ 78.904 \\
19b\textunderscore1\textunderscore10542 & 292.815320 & 36.294256 &
16.36 & M1 $\pm$ 1.19 & 2.060 & 0.038 $\pm$ 0.014 & --- & --- & --- \\
19b\textunderscore3\textunderscore03003 & 293.389460 & 36.815246 &
15.58 & M4 $\pm$ 1.00 & 1.480 & 0.015 $\pm$ 0.008 & --- & --- & --- \\
19b\textunderscore3\textunderscore08292 & 293.505170 & 36.904554 &
16.39 & K9 $\pm$ 1.05 & 28.710 & 0.028 $\pm$ 0.013 & 1.228 $\pm$ 3.669
& -3.199 $\pm$ 3.669 & 44.699 $\pm$ 4.699 \\
19b\textunderscore3\textunderscore09282 & 293.527870 & 36.925772 &
16.99 & M1 $\pm$ 1.06 & 7.680 & 0.028 $\pm$ 0.020 & --- & --- & --- \\
19b\textunderscore3\textunderscore12753 & 293.601230 & 36.745270 &
16.71 & M3 $\pm$ 1.01 & 2.450 & 0.037 $\pm$ 0.016 & --- & --- & --- \\
19c\textunderscore2\textunderscore05428 & 294.422680 & 36.416670 &
15.91 & K7 $\pm$ 1.00 & 20.680 & 0.115 $\pm$ 0.010 & 2.969 $\pm$ 2.681
& -2.105 $\pm$ 2.681 & 41.374 $\pm$ 3.151 \\
19c\textunderscore3\textunderscore01804 & 294.179440 & 36.758408 &
16.27 & M3 $\pm$ 1.00 & 7.700 & 0.036 $\pm$ 0.012 & --- & --- & --- \\
19c\textunderscore3\textunderscore03299 & 294.205130 & 36.760331 &
13.17 & M0 $\pm$ 1.00 & 5.620 & 0.016 $\pm$ 0.002 & 7.245 $\pm$ 2.438
& -18.289 $\pm$ 2.438 & 42.647 $\pm$ 2.796 \\
19c\textunderscore3\textunderscore04974 & 294.235120 & 36.750789 &
12.41 & M2 $\pm$ 1.64 & 1.450 & 0.010 $\pm$ 0.002 & -10.040 $\pm$
2.393 & 0.388 $\pm$ 2.393 & 21.279 $\pm$ 1.379 \\
19c\textunderscore3\textunderscore05921 & 294.252830 & 36.875064 &
15.37 & M4 $\pm$ 1.05 & 42.400 & 0.016 $\pm$ 0.006 & --- & --- & ---
\\ 19c\textunderscore3\textunderscore11273 & 294.354460 & 36.709272 &
15.75 & M3 $\pm$ 1.06 & 90.330 & 0.016 $\pm$ 0.008 & --- & --- & ---
\\ 19c\textunderscore4\textunderscore10571 & 293.851810 & 36.888130 &
15.92 & M4 $\pm$ 1.02 & 1.440 & 0.032 $\pm$ 0.011 & --- & --- & --- \\
19c\textunderscore4\textunderscore12623 & 293.864930 & 36.921243 &
16.54 & M3 $\pm$ 1.10 & 3.600 & 0.034 $\pm$ 0.016 & 3.934 $\pm$ 5.675
& -17.812 $\pm$ 5.675 & 88.260 $\pm$ 14.006 \\
19d\textunderscore1\textunderscore06078 & 294.046830 & 36.381880 &
15.14 & M1 $\pm$ 1.00 & 15.490 & 0.023 $\pm$ 0.006 & 2.722 $\pm$ 3.513
& 2.695 $\pm$ 3.513 & 15.304 $\pm$ 1.497 \\
19d\textunderscore1\textunderscore06603 & 294.036450 & 36.297312 &
16.21 & M2 $\pm$ 1.10 & 5.360 & 0.027 $\pm$ 0.012 & --- & --- & --- \\
19d\textunderscore1\textunderscore12693 & 293.923010 & 36.287590 &
16.00 & M4 $\pm$ 1.03 & 0.730 & 0.033 $\pm$ 0.010 & --- & --- & --- \\
19d\textunderscore2\textunderscore00740 & 294.453090 & 36.487943 &
15.54 & M3 $\pm$ 1.00 & 8.050 & 0.025 $\pm$ 0.008 & --- & --- & --- \\
19d\textunderscore3\textunderscore01140 & 294.441370 & 36.901497 &
15.64 & M4 $\pm$ 1.01 & 2.410 & 0.019 $\pm$ 0.008 & --- & --- & --- \\
19d\textunderscore3\textunderscore01271 & 294.445960 & 36.921308 &
16.38 & M4 $\pm$ 1.00 & 0.560 & 0.026 $\pm$ 0.013 & --- & --- & --- \\
19d\textunderscore3\textunderscore01937 & 294.455060 & 36.830460 &
12.63 & M3 $\pm$ 1.00 & 0.470 & 0.045 $\pm$ 0.002 & -17.728 $\pm$
2.344 & 1.854 $\pm$ 2.344 & 23.887 $\pm$ 1.434 \\
19d\textunderscore3\textunderscore02216 & 294.461610 & 36.796780 &
14.10 & M4 $\pm$ 1.00 & 2.010 & 0.023 $\pm$ 0.003 & 12.883 $\pm$ 4.176
& 5.053 $\pm$ 4.176 & 14.979 $\pm$ 1.616 \\
19d\textunderscore3\textunderscore03681 & 294.482160 & 36.713086 &
15.83 & M0 $\pm$ 1.00 & 17.480 & 0.019 $\pm$ 0.009 & --- & --- & ---
\\ 19d\textunderscore3\textunderscore05922 & 294.526470 & 36.803361 &
16.08 & M2 $\pm$ 1.12 & 23.060 & 0.029 $\pm$ 0.010 & -0.337 $\pm$
3.841 & 2.172 $\pm$ 3.841 & 11.717 $\pm$ 1.263 \\
19d\textunderscore3\textunderscore07286 & 294.550410 & 36.755946 &
15.38 & M0 $\pm$ 1.00 & 7.170 & 0.026 $\pm$ 0.007 & 1.442 $\pm$ 2.741
& 4.024 $\pm$ 2.741 & 25.633 $\pm$ 1.961 \\
19e\textunderscore1\textunderscore03204 & 292.733110 & 36.090610 &
16.79 & M1 $\pm$ 1.19 & 1.950 & 0.042 $\pm$ 0.018 & --- & --- & --- \\
19e\textunderscore2\textunderscore02880 & 293.130710 & 36.239925 &
15.18 & M4 $\pm$ 1.00 & 50.020 & 0.013 $\pm$ 0.006 & 3.037 $\pm$ 3.696
& 3.874 $\pm$ 3.696 & 8.848 $\pm$ 0.920 \\
19e\textunderscore2\textunderscore09200 & 293.301440 & 36.154486 &
15.22 & M4 $\pm$ 1.00 & 1.350 & 0.018 $\pm$ 0.006 & --- & --- & --- \\
19e\textunderscore3\textunderscore09622 & 293.199860 & 36.614838 &
16.46 & M4 $\pm$ 1.00 & 1.620 & 0.025 $\pm$ 0.014 & --- & --- & --- \\
19f\textunderscore3\textunderscore08798 & 293.508310 & 36.571411 &
15.98 & M4 $\pm$ 1.00 & 65.390 & 0.026 $\pm$ 0.010 & 12.862 $\pm$
4.506 & 13.205 $\pm$ 4.506 & 39.388 $\pm$ 4.858 \\
19f\textunderscore3\textunderscore12800 & 293.593410 & 36.489227 &
15.97 & M3 $\pm$ 1.18 & 3.890 & 0.018 $\pm$ 0.009 & --- & --- & --- \\
19f\textunderscore4\textunderscore06780 & 292.808650 & 36.620585 &
17.02 & M0 $\pm$ 1.76 & 2.620 & 0.048 $\pm$ 0.022 & --- & --- & --- \\
19g\textunderscore1\textunderscore04348 & 293.805310 & 36.117012 &
16.34 & M0 $\pm$ 1.06 & 1.420 & 0.042 $\pm$ 0.014 & 3.106 $\pm$ 3.109
& 1.070 $\pm$ 3.109 & 32.632 $\pm$ 2.962 \\
19g\textunderscore1\textunderscore05313 & 293.787750 & 36.249911 &
16.76 & M4 $\pm$ 1.00 & 34.220 & 0.041 $\pm$ 0.017 & --- & --- & ---
\\ 19g\textunderscore1\textunderscore10773 & 293.683380 & 36.080351 &
16.16 & M4 $\pm$ 1.08 & 5.910 & 0.023 $\pm$ 0.011 & --- & --- & --- \\
19g\textunderscore2\textunderscore03649 & 294.179550 & 36.224328 &
16.47 & M3 $\pm$ 1.00 & 9.420 & 0.053 $\pm$ 0.015 & --- & --- & --- \\
19g\textunderscore3\textunderscore06870 & 294.268510 & 36.556675 &
15.62 & M4 $\pm$ 1.06 & 13.000 & 0.014 $\pm$ 0.008 & -8.732 $\pm$
3.748 & -1.280 $\pm$ 3.748 & 18.177 $\pm$ 1.921 \\
19g\textunderscore4\textunderscore00989 & 293.744820 & 36.504677 &
16.20 & M4 $\pm$ 1.20 & 40.050 & 0.025 $\pm$ 0.012 & --- & --- & ---
\\ 19h\textunderscore3\textunderscore16170 & 294.703050 & 36.655522 &
16.99 & M0 $\pm$ 1.31 & 7.560 & 0.052 $\pm$ 0.018 & -5.934 $\pm$ 3.846
& -12.512 $\pm$ 3.846 & 309.087 $\pm$ 34.741 \\ \hline
\end{tabular}
\begin{tabular}{p{\textwidth}} $^a$ Coordinates in SDSS (epoch J2000)
if available\\ $^b$ WTS average UKIDSS system $J$ band magnitude MKO
system\\ $^c$ Spectral type estimate as determined by method described
in Section \ref{sec:Typing} with error determined by standard
deviation in colour-type space and minimum uncertainty of $\pm$1 (see
Figure \ref{fig:hist})\\ $^d$ Period in days as determined by
maximum peak in Lomb-Scargle periodogram\\ $^e$ Amplitude in $J$ band
magnitude\\ $^f$ Proper motion in masyr$^{-1}$ from SDSS\\ $^g$
Tangential velocity in kms$^{-1}$
\end{tabular}\footnotesize
\caption{A table of the 68 variable M dwarfs found. Included are stars
outside the M dwarf spectral type range but with H-K $>$ 0.175
suggesting a redder, late type star that the spectral typing procedure
erroneous assigned an earlier spectral type, and have been retained
for maximum inclusivity. Spectroscopic follow up is required to
absolutely confirm the nature of these stars, and any subsequent
exclusion would not effect the conclusions we draw.}
\label{tab:var}
\end{table*}


\begin{figure*} \hbox{\includegraphics[scale=0.9]{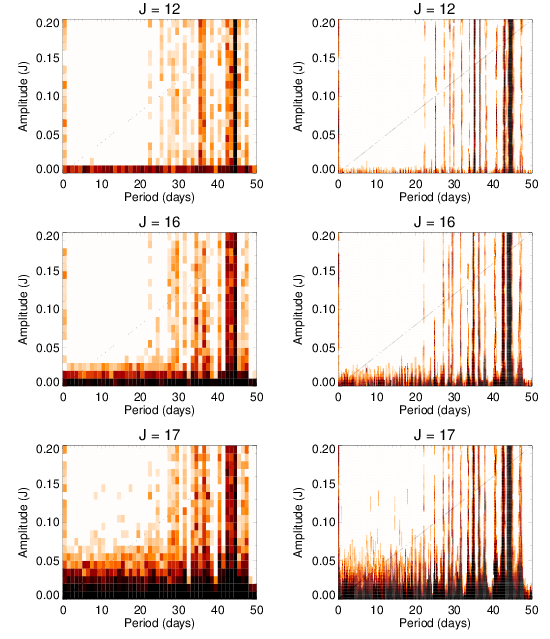}}
\caption{Fractions of injected signal detected for lightcurve
amplitude as a function of period. White presents a retrieval fraction
of 1 (i.e. a signal corresponding to the period with a probability
$>$.99 of a real signal not caused by noise is retrieved in every
simulation) and black a fraction of 0 (i.e. no significant signal
corresponding to the injected period is found and noise dominates in
all simulations). The plots on the left show the output binned at 1
day period intervals and 0.01 $J$ band magnitude amplitude intervals
with an average of 1666.7 data per bin, and on the right at 0.25 day
and 0.005 $J$ band magnitude intervals respectively with an average of
4.2 data per bin. A window of high-retrievability remains for the
brightest targets for $J$ = 12 to $J$ = 15 with detectability
decreasing at $J$ $>$ 16. The diagrams on the right show a very strong
dependency on period due to aliasing effects with periods of
observation. Please see online article for a colour version of this
figure.}
\label{fig:mcplots}
\end{figure*}

\begin{figure*}
\hbox{\includegraphics[angle=270,scale=0.7]{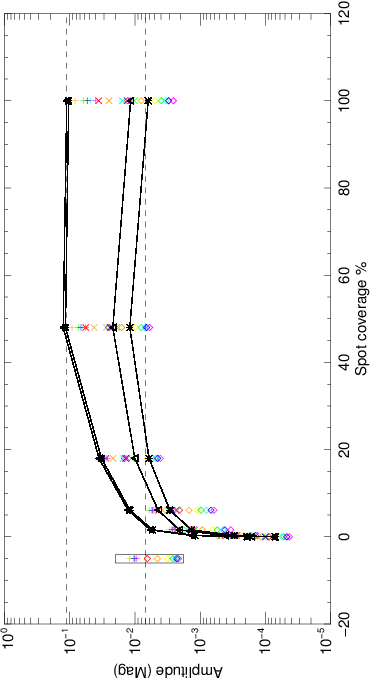}}
\caption{Output from the DoTS code of peak to peak amplitudes for
simulated light curves against the spot coverage percentage. Coloured
diamonds, crosses and pluses indicate $J$ band scenarios for $T_s =
T_p - 250$K, $T_s=0.8T_p$ and $T_s=0.6T_p$ with each colour
representing a photosphere temperate from 2250K (red) to 3750K
(magenta) in 250K steps. The joined stars indicate the analogous
maximum and minimum $I$ band amplitudes in each case, and the triangle
the analogous $V$ band amplitudes respectively. The dashed lines
represent the range of amplitudes detected in our sample of variable
stars; the low limit imposed by the noise limitation in WTS. The boxed
points show the amplitudes found by having one large spot group
located on the stars equator. The $I$ and $V$ band amplitudes for the
less spotted cases are larger, indicating that we are sampling stars
with activities greater then $\sim10$ per cent and that surveys in the
$J$ band such as WTS may not be sensitive to lower amplitude
variability in the less spotted cases than surveys at shorter
wavelengths. The amplitude downturn between the two most heavily
spotted models is due to the increase in homogeneity as the the spot
coverages tends towards totality, such that the temperature
differentials on the photosphere become dominated by the lesser
contrast between spot umbrae and penumbrae rather than the greater
contrast between spotted and unspotted areas. Please see online
article for a colour version of this figure.}
\label{fig:dots}
\end{figure*}

\begin{figure*} \hbox{\includegraphics[scale=0.92,clip]{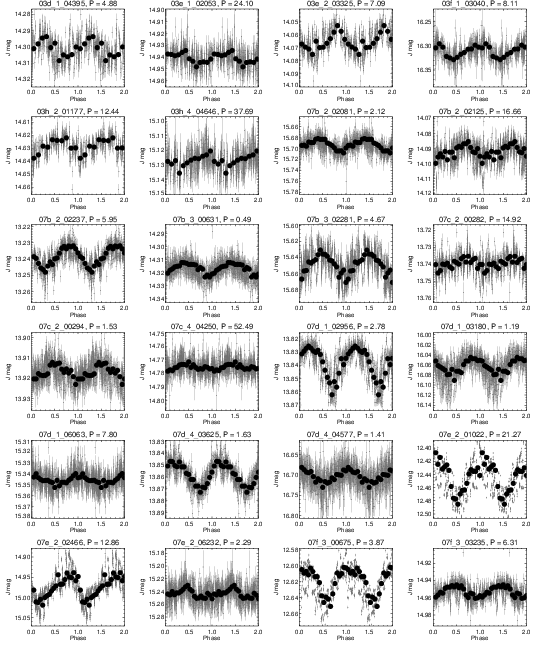}}
\caption{Folded light curves for the variable stars with estimated
spectral times in the 03, 07 and 19 hour fields. Light curves are
labeled with the name of the star and the period from from the
periodogram.}
\label{fig:lc1}
\end{figure*}

\begin{figure*} \hbox{\includegraphics[scale=0.92,clip]{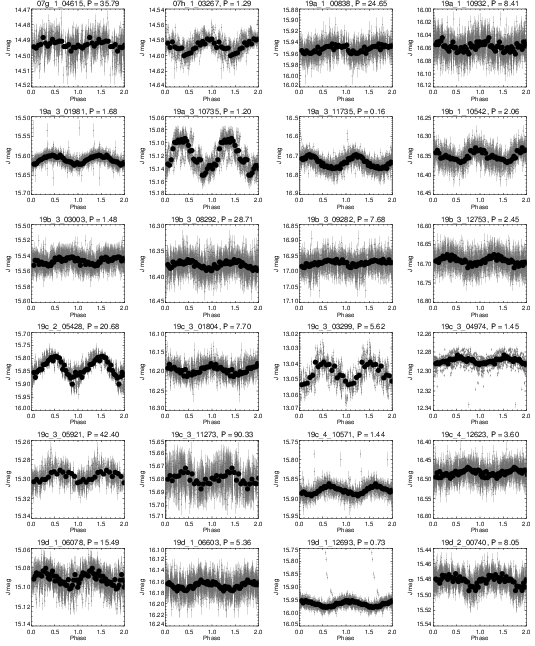}}
\caption{As in Figure \ref{fig:lc1}.}
\label{fig:lc2}
\end{figure*}

\begin{figure*} \hbox{\includegraphics[trim=0cm 4cm 0cm
0cm,scale=0.92,clip]{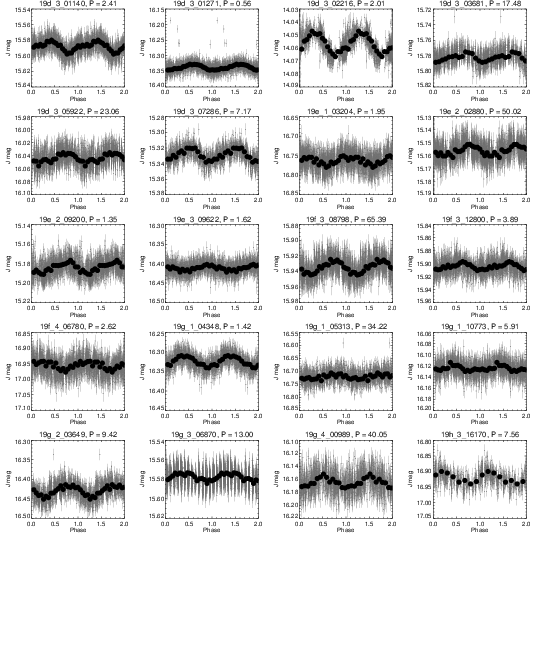}}
\caption{As in Figure \ref{fig:lc1}.}
\label{fig:lc3}
\end{figure*}

\begin{figure*} \hbox{\includegraphics[scale=0.7, clip]{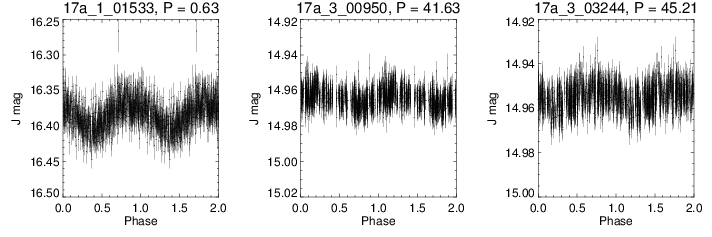}}
\caption{As in Figure \ref{fig:example}. These light curves are the
periodically variable stars found in the 17hr field for which there
are no observations in the SDSS and therefore no spectral type
estimation were made.}
\label{fig:17hrp}
\end{figure*}


\begin{thebibliography}{53} \expandafter\ifx\csname
natexlab\endcsname\relax\def\natexlab#1{#1}\fi

\bibitem[{{Baraffe} \& {Chabrier}(1996)}]{Baraffe1996} {Baraffe},
I. \& {Chabrier}, G. 1996, \apjl, 461, L51+

\bibitem[{Barnes {et~al.}(2002)Barnes, James, \& Cameron}]{Barnes2003}
Barnes, J., James, D., \& Cameron, A. 2002, Astronomische Nachrichten,
323, 333

\bibitem[{{Barnes} {et~al.}(2011){Barnes}, {Jeffers}, \&
{Jones}}]{Barnes2011} {Barnes}, J.~R., {Jeffers}, S.~V., \& {Jones},
H.~R.~A. 2011, \mnras, 412, 1599

\bibitem[{{Benedict} {et~al.}(1998){Benedict}, {McArthur}, {Nelan},
{Story}, {Whipple}, {Shelus}, {Jefferys}, {Hemenway}, {Franz},
{Wasserman}, {Duncombe}, {van Altena}, \& {Fredrick}}]{Benedict1998}
{Benedict}, G.~F., {McArthur}, B., {Nelan}, E., {et~al.} 1998, \aj,
116, 429

\bibitem[{{Berta} {et~al.}(2012){Berta}, {Irwin}, {Charbonneau},
{Burke}, \& {Falco}}]{Berta2012} {Berta}, Z.~K., {Irwin}, J.,
{Charbonneau}, D., {Burke}, C.~J., \& {Falco}, E.~E. 2012, ArXiv
e-prints

\bibitem[{{Bochanski} {et~al.}(2011){Bochanski}, {Hawley}, \&
{West}}]{Bochanski2011} {Bochanski}, J.~J., {Hawley}, S.~L., \&
{West}, A.~A. 2011, \aj, 141, 98

\bibitem[{{Borucki} {et~al.}(2011){Borucki}, {Koch}, {Basri},
{Batalha}, {Brown}, {Bryson}, {Caldwell}, {Christensen-Dalsgaard},
{Cochran}, {DeVore}, {Dunham}, {Gautier}, {Geary}, {Gilliland},
{Gould}, {Howell}, {Jenkins}, {Latham}, {Lissauer}, {Marcy}, {Rowe},
{Sasselov}, {Boss}, {Charbonneau}, {Ciardi}, {Doyle}, {Dupree},
{Ford}, {Fortney}, {Holman}, {Seager}, {Steffen}, {Tarter}, {Welsh},
{Allen}, {Buchhave}, {Christiansen}, {Clarke}, {Das}, {D{\'e}sert},
{Endl}, {Fabrycky}, {Fressin}, {Haas}, {Horch}, {Howard}, {Isaacson},
{Kjeldsen}, {Kolodziejczak}, {Kulesa}, {Li}, {Lucas}, {Machalek},
{McCarthy}, {MacQueen}, {Meibom}, {Miquel}, {Prsa}, {Quinn},
{Quintana}, {Ragozzine}, {Sherry}, {Shporer}, {Tenenbaum}, {Torres},
{Twicken}, {Van Cleve}, {Walkowicz}, {Witteborn}, \&
{Still}}]{Borucki2011} {Borucki}, W.~J., {Koch}, D.~G., {Basri}, G.,
{et~al.} 2011, \apj, 736, 19

\bibitem[{{Brown} {et~al.}(2008){Brown}, {Browning}, {Brun}, {Miesch},
\& {Toomre}}]{Brown2008} {Brown}, B.~P., {Browning}, M.~K., {Brun},
A.~S., {Miesch}, M.~S., \& {Toomre}, J. 2008, \apj, 689, 1354

\bibitem[{{Browning} {et~al.}(2010){Browning}, {Basri}, {Marcy},
{West}, \& {Zhang}}]{Browning2010} {Browning}, M.~K., {Basri}, G.,
{Marcy}, G.~W., {West}, A.~A., \& {Zhang}, J.  2010, \aj, 139, 504

\bibitem[{{Chabrier} \& {K{\"u}ker}(2006)}]{Chabrier2006} {Chabrier},
G. \& {K{\"u}ker}, M. 2006, \aap, 446, 1027

\bibitem[{{Charbonneau} {et~al.}(2009){Charbonneau}, {Berta}, {Irwin},
{Burke}, {Nutzman}, {Buchhave}, {Lovis}, {Bonfils}, {Latham}, {Udry},
{Murray-Clay}, {Holman}, {Falco}, {Winn}, {Queloz}, {Pepe}, {Mayor},
{Delfosse}, \& {Forveille}}]{Charbonneau2009} {Charbonneau}, D.,
{Berta}, Z.~K., {Irwin}, J., {et~al.} 2009, \nat, 462, 891

\bibitem[{{Ciardi} {et~al.}(2011){Ciardi}, {von Braun}, {Bryden}, {van
Eyken}, {Howell}, {Kane}, {Plavchan}, {Ram{\'{\i}}rez}, \&
{Stauffer}}]{Ciardi2011} {Ciardi}, D.~R., {von Braun}, K., {Bryden},
G., {et~al.} 2011, \aj, 141, 108

\bibitem[{{Collier Cameron}(1997)}]{Cameron1997} {Collier Cameron},
A. 1997, \mnras, 287, 556

\bibitem[{{Coughlin} {et~al.}(2008){Coughlin}, {Stringfellow},
{Becker}, {L{\'o}pez-Morales}, {Mezzalira}, \&
{Krajci}}]{Coughlin2008} {Coughlin}, J.~L., {Stringfellow}, G.~S.,
{Becker}, A.~C., {et~al.} 2008, \apjl, 689, L149

\bibitem[{Covey {et~al.}(2007)}]{Covey2007} Covey, K.~R. {et~al.}
2007, Astron. J., 134, 2398

\bibitem[{{Davenport} {et~al.}(2012){Davenport}, {Becker}, {Kowalski},
{Hawley}, {Schmidt}, {Hilton}, {Sesar}, \& {Cutri}}]{Davenport2012}
{Davenport}, J.~R.~A., {Becker}, A.~C., {Kowalski}, A.~F., {et~al.}
2012, \apj, 748, 58

\bibitem[{{Frasca} {et~al.}(2009){Frasca}, {Covino}, {Spezzi},
{Alcal{\'a}}, {Marilli}, {F{\.z}r{\'e}sz}, \& {Gandolfi}}]{Frasca2009}
{Frasca}, A., {Covino}, E., {Spezzi}, L., {et~al.} 2009, \aap, 508,
1313

\bibitem[{{Fressin} {et~al.}(2011){Fressin}, {Torres}, {Rowe},
{Charbonneau}, {Rogers}, {Ballard}, {Batalha}, {Borucki}, {Bryson},
{Buchhave}, {Ciardi}, {Desert}, {Dressing}, {Fabrycky}, {Ford},
{Gautier}, {Henze}, {Holman}, {Howard}, {Howell}, {Jenkins}, {Koch},
{Latham}, {Lissauer}, {Marcy}, {Quinn}, {Ragozzine}, {Sasselov},
{Seager}, {Barclay}, {Mullally}, {Seader}, {Still}, {Twicken},
{Thompson}, \& {Uddin}}]{Fressin2011} {Fressin}, F., {Torres}, G.,
{Rowe}, J.~F., {et~al.} 2011, ArXiv e-prints

\bibitem[{{Gillon} {et~al.}(2007){Gillon}, {Pont}, {Demory},
{Mallmann}, {Mayor}, {Mazeh}, {Queloz}, {Shporer}, {Udry}, \&
{Vuissoz}}]{Gillon2007} {Gillon}, M., {Pont}, F., {Demory}, B.-O.,
{et~al.} 2007, \aap, 472, L13

\bibitem[{{Granzer} {et~al.}(2000){Granzer}, {Sch{\"u}ssler},
{Caligari}, \& {Strassmeier}}]{Granzer2000} {Granzer}, T.,
{Sch{\"u}ssler}, M., {Caligari}, P., \& {Strassmeier}, K.~G.  2000,
\aap, 355, 1087

\bibitem[{{Hartman} {et~al.}(2011){Hartman}, {Bakos}, {Noyes}, {Sip{\H
o}cz}, {Kov{\'a}cs}, {Mazeh}, {Shporer}, \& {P{\'a}l}}]{Hartman2011}
{Hartman}, J.~D., {Bakos}, G.~{\'A}., {Noyes}, R.~W., {et~al.} 2011,
\aj, 141, 166

\bibitem[{Hewett {et~al.}(2006)Hewett, Warren, Leggett, \&
Hodgkin}]{Hewett2006} Hewett, P.~C., Warren, S.~J., Leggett, S.~K., \&
Hodgkin, S.~T. 2006, Mon. Not.  Roy. Astron. Soc., 367, 454

\bibitem[{{Horne} \& {Baliunas}(1986)}]{Horne1986} {Horne}, J.~H. \&
{Baliunas}, S.~L. 1986, \apj, 302, 757

\bibitem[{{Irwin} {et~al.}(2011){Irwin}, {Berta}, {Burke},
{Charbonneau}, {Nutzman}, {West}, \& {Falco}}]{Irwin2011} {Irwin}, J.,
{Berta}, Z.~K., {Burke}, C.~J., {et~al.} 2011, \apj, 727, 56

\bibitem[{{Irwin} {et~al.}(2007){Irwin}, {Irwin}, {Aigrain},
{Hodgkin}, {Hebb}, \& {Moraux}}]{Irwin2007} {Irwin}, J., {Irwin}, M.,
{Aigrain}, S., {et~al.} 2007, \mnras, 375, 1449

\bibitem[{{Jackson} {et~al.}(2009){Jackson}, {Jeffries}, \&
{Maxted}}]{Jackson2009} {Jackson}, R.~J., {Jeffries}, R.~D., \&
{Maxted}, P.~F.~L. 2009, \mnras, 399, L89

\bibitem[{{Jeffers} {et~al.}(2007){Jeffers}, {Donati}, \& {Collier
Cameron}}]{Jeffers2007} {Jeffers}, S.~V., {Donati}, J.-F., \& {Collier
Cameron}, A. 2007, \mnras, 375, 567

\bibitem[{{Johnson} {et~al.}(2012){Johnson}, {Gazak}, {Apps},
{Muirhead}, {Crepp}, {Crossfield}, {Boyajian}, {von Braun},
{Rojas-Ayala}, {Howard}, {Covey}, {Schlawin}, {Hamren}, {Morton},
{Marcy}, \& {Lloyd}}]{Johnson2012} {Johnson}, J.~A., {Gazak}, J.~Z.,
{Apps}, K., {et~al.} 2012, \aj, 143, 111

\bibitem[{{Kiraga} \& {St\c{e}pie\'{n}}(2007)}]{Kiraga2007} {Kiraga},
M. \& {St\c{e}pie\'{n}}, K. 2007, \actaa, 57, 149

\bibitem[{{{Kov\'{a}cs}, G. and {Hodgkin}, S. and {Sip\H{o}cz}, B and
{Pinfield}, D., and others}(2012)}]{Kovacs2012} {{Kov\'{a}cs}, G. and
{Hodgkin}, S. and {Sip\H{o}cz}, B and {Pinfield}, D., and
others}. 2012, \inprep

\bibitem[{{Lomb}(1976)}]{Lomb1976} {Lomb}, N.~R. 1976, \apss, 39, 447

\bibitem[{{Mamajek}(2011)}]{Mamajek2011} {Mamajek}, E. 2011, {A Modern
Mean Stellar Color and Effective Temperatures (Teff) Sequence for
O9V-Y0V Dwarf Stars}, {See
http://www.pas.rochester.edu/$\sim$emamajek/EEM\_dwarf
\_UBVIJHK\_colors\_Teff.dat [Online; accessed 09-August-2012]}

\bibitem[{{Messina} {et~al.}(2011){Messina}, {Desidera}, {Lanzafame},
{Turatto}, \& {Guinan}}]{Messina2011} {Messina}, S., {Desidera}, S.,
{Lanzafame}, A.~C., {Turatto}, M., \& {Guinan}, E.~F. 2011, \aap, 532,
A10+

\bibitem[{{Messina} {et~al.}(2003){Messina}, {Pizzolato}, {Guinan}, \&
{Rodon{\`o}}}]{Messina2003} {Messina}, S., {Pizzolato}, N., {Guinan},
E.~F., \& {Rodon{\`o}}, M. 2003, \aap, 410, 671

\bibitem[{{Miller} {et~al.}(2008){Miller}, {Irwin}, {Aigrain},
{Hodgkin}, \& {Hebb}}]{Miller2008} {Miller}, A.~A., {Irwin}, J.,
{Aigrain}, S., {Hodgkin}, S., \& {Hebb}, L. 2008, \mnras, 387, 349

\bibitem[{{Miller}(1966)}]{Miller1966} {Miller}, W.~J. 1966, Ricerche
Astronomiche, 7, 217

\bibitem[{{Mohanty} \& {Basri}(2003)}]{Mohanty2003} {Mohanty}, S. \&
{Basri}, G. 2003, \apj, 583, 451

\bibitem[{{Morales} {et~al.}(2010){Morales}, {Gallardo}, {Ribas},
{Jordi}, {Baraffe}, \& {Chabrier}}]{Morales2010} {Morales}, J.~C.,
{Gallardo}, J., {Ribas}, I., {et~al.} 2010, \apj, 718, 502

\bibitem[{{Moreno-Insertis} {et~al.}(1992){Moreno-Insertis},
{Sch\"{u}ssler}, \& {Ferriz-Mas}}]{Moreno1992} {Moreno-Insertis}, F.,
{Sch\"{u}ssler}, M., \& {Ferriz-Mas}, A. 1992, \aap, 264, 686

\bibitem[{{Morin} {et~al.}(2008){Morin}, {Donati}, {Petit},
{Delfosse}, {Forveille}, {Albert}, {Auri{\`e}re}, {Cabanac},
{Dintrans}, {Fares}, {Gastine}, {Jardine}, {Ligni{\`e}res}, {Paletou},
{Ramirez Velez}, \& {Th{\'e}ado}}]{Morin2008} {Morin}, J., {Donati},
J.-F., {Petit}, P., {et~al.} 2008, \mnras, 390, 567

\bibitem[{{Morin} {et~al.}(2010){Morin}, {Donati}, {Petit},
{Delfosse}, {Forveille}, \& {Jardine}}]{Morin2010} {Morin}, J.,
{Donati}, J.-F., {Petit}, P., {et~al.} 2010, \mnras, 407, 2269

\bibitem[{{Muirhead} {et~al.}(2012){Muirhead}, {Johnson}, {Apps},
{Carter}, {Morton}, {Fabrycky}, {Pineda}, {Bottom}, {Rojas-Ayala},
{Schlawin}, {Hamren}, {Covey}, {Crepp}, {Stassun}, {Pepper}, {Hebb},
{Kirby}, {Howard}, {Isaacson}, {Marcy}, {Levitan}, {Diaz-Santos},
{Armus}, \& {Lloyd}}]{Muirhead2012} {Muirhead}, P.~S., {Johnson},
J.~A., {Apps}, K., {et~al.} 2012, ArXiv e-prints

\bibitem[{{Nefs} {et~al.}(2012){Nefs}, {Birkby}, {Snellen}, {Hodgkin},
{Pinfield}, {Sipocz}, {Kovacs}, {Mislis}, {Saglia}, {Koppenhofer},
{Cruz}, {Barrado}, {Martin}, {Goulding}, {Stoev}, {Zendejas}, {del
Burgo}, {Cappetta}, \& {Pavlenko}}]{Nefs2012} {Nefs}, S.~V., {Birkby},
J.~L., {Snellen}, I.~A.~G., {et~al.} 2012, ArXiv e-prints

\bibitem[{{Plavchan} {et~al.}(2008){Plavchan}, {Jura}, {Kirkpatrick},
{Cutri}, \& {Gallagher}}]{Plavchan2008} {Plavchan}, P., {Jura}, M.,
{Kirkpatrick}, J.~D., {Cutri}, R.~M., \& {Gallagher}, S.~C. 2008,
\apjs, 175, 191

\bibitem[{{Reiners}(2007)}]{Reiners2007} {Reiners}, A. 2007, \aap,
467, 259

\bibitem[{{Reiners} \& {Basri}(2008)}]{Reiners2008} {Reiners}, A. \&
{Basri}, G. 2008, \apj, 684, 1390

\bibitem[{{Rockenfeller} {et~al.}(2006){Rockenfeller}, {Bailer-Jones},
\& {Mundt}}]{Rockenfeller2006} {Rockenfeller}, B., {Bailer-Jones},
C.~A.~L., \& {Mundt}, R. 2006, \aap, 448, 1111

\bibitem[{{Scargle}(1982)}]{Scargle1982} {Scargle}, J.~D. 1982, \apj,
263, 835

\bibitem[{{Schuessler} \& {Solanki}(1992)}]{Schuessler1992}
{Schuessler}, M. \& {Solanki}, S.~K. 1992, \aap, 264, L13

\bibitem[{{Tofflemire} {et~al.}(2012){Tofflemire}, {Wisniewski},
{Kowalski}, {Schmidt}, {Kundurthy}, {Hilton}, {Holtzman}, \&
{Hawley}}]{Tofflemire2012} {Tofflemire}, B.~M., {Wisniewski}, J.~P.,
{Kowalski}, A.~F., {et~al.} 2012, \aj, 143, 12

\bibitem[{{Vogt} {et~al.}(2010){Vogt}, {Butler}, {Rivera},
{Haghighipour}, {Henry}, \& {Williamson}}]{Vogt2010} {Vogt}, S.~S.,
{Butler}, R.~P., {Rivera}, E.~J., {et~al.} 2010, \apj, 723, 954

\bibitem[{{West} {et~al.}(2008){West}, {Hawley}, {Bochanski}, {Covey},
{Reid}, {Dhital}, {Hilton}, \& {Masuda}}]{West2008} {West}, A.~A.,
{Hawley}, S.~L., {Bochanski}, J.~J., {et~al.} 2008, \aj, 135, 785

\bibitem[{{West} {et~al.}(2011){West}, {Morgan}, {Bochanski},
{Andersen}, {Bell}, {Kowalski}, {Davenport}, {Hawley}, {Schmidt},
{Bernat}, {Hilton}, {Muirhead}, {Covey}, {Rojas-Ayala}, {Schlawin},
{Gooding}, {Schluns}, {Dhital}, {Pineda}, \& {Jones}}]{West2011}
{West}, A.~A., {Morgan}, D.~P., {Bochanski}, J.~J., {et~al.} 2011,
\aj, 141, 97

\end{thebibliography}
\end{document}